\documentclass[letter]{aa} 

\usepackage{natbib}
\bibpunct{(}{)}{;}{a}{}{,} 

\usepackage{graphicx}
\usepackage{subfigure}
\usepackage{txfonts}
\usepackage[colorlinks=True,allcolors=blue]{hyperref}
\begin{document}

   \title{Evidence of extended [CII] and dust emission in local dwarf galaxies}
   \authorrunning{M. Romano et al.}
   
\author{M. Romano\thanks{E-mail: michael.romano@ncbj.gov.pl}\inst{1,2}
\and
D. Donevski\inst{1,3,4}
\and
Junais\inst{1}
\and
A. Nanni\inst{1,5}
\and
M. Ginolfi\inst{6}
\and
G. C. Jones\inst{7}
\and
I. Shivaei\inst{8}
\and
G. Lorenzon\inst{1}
\and
M. Hamed\inst{1}
\and
D. Salak\inst{9,10}
\and
P. Sawant\inst{1}
}

\institute{
National Centre for Nuclear Research, ul. Pasteura 7, 02-093 Warsaw, Poland
\and
INAF - Osservatorio Astronomico di Padova, Vicolo dell'Osservatorio 5, I-35122, Padova, Italy
\and
SISSA, Via Bonomea 265, Trieste, Italy
\and
IFPU - Institute for fundamental physics of the Universe, Via Beirut 2, 34014 Trieste, Italy
\and
INAF - Osservatorio astronomico d'Abruzzo, Via Maggini SNC, 64100, Teramo, Italy
\and
Dipartimento di Fisica e Astronomia, Università di Firenze, Via G. Sansone 1, 50019, Sesto Fiorentino (Firenze), Italy
\and
Department of Physics, University of Oxford, Denys Wilkinson Building, Keble Road, Oxford OX1 3RH, UK
\and
Centro de Astrobiolog\'{i}a (CAB), CSIC-INTA, Ctra. de Ajalvir km 4, Torrej\'{o}n de Ardoz, E-28850, Madrid, Spain
\and
Institute for the Advancement of Higher Education, Hokkaido University, Kita 17 Nishi 8, Kita-ku, Sapporo, Hokkaido 060-0817,
Japan
\and
Department of Cosmosciences, Graduate School of Science,
Hokkaido University, Kita 10 Nishi 8, Kita-ku, Sapporo, Hokkaido
060-0810, Japan
}

\abstract
{}
{The evolution of dwarf galaxies is dramatically affected by gaseous and dusty outflows, which can easily deprive their interstellar medium of the material needed for the formation of new stars, simultaneously enriching their surrounding circumgalactic medium (CGM). In this letter, we present the first evidence of extended [CII]~158~$\mu$m line and dust continuum emission in local dwarf galaxies hosting star-formation-driven outflows.}
{By stacking the [CII], far-infrared, and near-UV (NUV) emission obtained from \textit{Herschel} and GALEX data, we derived the average radial profiles, and compared the spatial extension of gas, dust, and stellar activity in dwarf galaxies.}
{We find that [CII] and dust emissions are comparable to each other, and more extended than the NUV continuum. The [CII] size is in agreement with that measured for $z>4$ star-forming galaxies, suggesting that similar mechanisms could be at the origin of the observed atomic carbon reservoir around local and high-$z$ sources. The cold dust follows the [CII] emission, going beyond the stellar continuum as opposed to what is typically observed in the early Universe where measurements can be affected by the poor sensitivity and faintness of dust emission in the CGM of high-$z$ galaxies.}
{We attribute the extended [CII] and dust continuum emission to the presence of galactic outflows. As local dwarf galaxies are considered analogs of primordial sources, we   expect that comparable feedback processes can be at the origin of the observed [CII] halos at $z>4$, dominating over other possible formation mechanisms.} 
 
  
\keywords{Galaxies: dwarf - Galaxies: evolution - Galaxies: ISM - Galaxies: starburst - ISM: jets and outflows}

\maketitle
%

\section{Introduction}
The evolution of galaxies is regulated by the tangled interplay between gas, stars, and dust in their interstellar medium (ISM). This picture is further complicated by a continuous exchange of material with the surrounding circumgalactic medium (CGM) that can be driven by merging activity, inflow, and outflow of gas, giving rise to what is known as the   baryon cycle (see, e.g., \citealt{Tumlinson17,Peroux20} for a review). Typically, the CGM can extend up to several kiloparsec, hosting a large reservoir of cosmic web-accreted gas and/or processed material ejected from galaxies that can be used as fuel for future star formation. 

Gas can be expelled from the ISM and injected into the CGM (or even farther into the intergalactic medium, IGM) through high-velocity winds produced by stellar feedback (e.g., \citealt{Gallerani18,Ginolfi20b,Romano23}) and active galactic nuclei (AGNs; e.g., \citealt{Cicone15,Rupke17,Jones23}). The ejected material can be made of dust, and ionized, atomic, and molecular gas that can be explored with different tracers and methods, including indirect observations of UV absorption lines along quasar lines of sight (e.g., \citealt{Werk16,Guo20}), or direct detections of emission lines such as HI 21 cm, Ly$\alpha$, [CII] 158~$\mu$m, or CO transitions (e.g., \citealt{Fujimoto20,Sanderson21,Ianjamasimanana22,Jones23}). 

Particularly, the [CII] line has gained a prominent role in the field of galaxy evolution, as it is a major coolant of the ISM and can provide precious information on the gas kinematics and morphology, star formation activity, stellar feedback, and CGM enrichment (e.g., \citealt{Ginolfi20a,Jones21,Romano21}). Due to its low ionization potential (i.e., 11.3 eV, as compared to the 13.6 eV of neutral hydrogen), [CII] can trace different gas phases (e.g., \citealt{Pineda13,Zanella18,Heintz21}), although many studies suggest that the bulk of its emission originates from the cold neutral gas (e.g., \citealt{Diaz-Santos17,Cormier19}). Recently, ALMA observations have revealed the existence of diffuse [CII] components (sometimes referred to  as [CII] halos) around $z>4$ galaxies, that are more extended than the underlying dust and stellar emission traced by the rest-frame UV and far-infrared (FIR) continuum (e.g., \citealt{Carniani18,Fujimoto19,Fujimoto20,Ginolfi20a,Herrera-Camus21,Akins22,Fudamoto22,Lambert23,Mitsuhashi23}). Such a wide gas distribution can be ascribed to different scenarios (e.g., satellites around the central galaxy, cold gas streams; \citealt{Fujimoto19,DiCesare24}), although it is often linked to the presence of star-formation-driven winds that can drag the gas outside of the galaxies enriching the surrounding CGM (e.g., \citealt{Hopkins12,Ginolfi20b,Pizzati20,Herrera-Camus21}). On the other hand, the spatial distribution of [CII], as compared to the UV and FIR emission, has been poorly investigated in the local Universe, with only a few individual sources showing hints of extended [CII] emission (e.g., \citealt{Madden93,Veilleux21}). 

In this letter, we report the first evidence of extended [CII] and dust emission around nearby dwarf galaxies. We used the sample of dwarf sources in \cite{Romano23}, where we found signatures of atomic\footnote{In \cite{Romano23}, we  assume that $\sim70\%$ of the [CII] emission originates from the atomic phase (e.g., \citealt{Cormier19}).} outflowing gas through \textit{Herschel}/PACS [CII] observations. In that work we found clues for outflow-induced CGM enrichment, and for extended [CII] emission as compared to the UV continuum. This motivated us to undertake a more in-depth investigation of the spatial distribution of [CII] in those galaxies, along with its origin and link to [CII] halos in more distant sources. 

The sample of dwarf galaxies is described in Sect. \ref{sec:sample}. In Sect. \ref{sec:method} we report the method used to stack the UV, FIR, and [CII] emission of the galaxies, and to retrieve the corresponding radial profiles. We present and discuss our results in Sect. \ref{sec:results}, and finally provide our conclusions in Sect. \ref{sec:conclusions}. We adopt a $\Lambda$CDM cosmology with $H_0 = 70~\mathrm{km~s^{-1}~Mpc^{-1}}$, $\mathrm{\Omega_m = 0.3,}$ and $\mathrm{\Omega_\Lambda = 0.7}$. At the mean redshift of the sample (i.e., $z_\mathrm{mean}=0.012$), 1 arcsecond corresponds to 0.2~kpc.

\section{Sample and observations}\label{sec:sample}
We analyzed the sample of 29 dwarf sources in \cite{Romano23}. Those targets were drawn from the Dwarf Galaxy Survey (DGS; \citealt{Madden13,Madden14}) that observed a collection of 48 low-metallicity ($7.14<\mathrm{12+log(O/H)<8.43}$) local dwarf galaxies (at distances $<200~$Mpc) in the FIR with the PACS \citep{Poglitsch10} and SPIRE \citep{Griffin10} instruments mounted on the \textit{Herschel} Space Observatory. All the galaxies benefit from a wide photometric coverage, going from the UV to the submillimeter \citep{Madden13}. Of particular relevance for this work are the observed UV and FIR continuum emissions. We used the near-UV (NUV; $\lambda_{\mathrm{eff}}\sim2271~\AA$) and 160~$\mu$m emission detected by the Galaxy Evolution Explorer (GALEX; \citealt{Martin05}) and \textit{Herschel} telescopes as proxies for recent star formation from young stars and dust continuum, respectively, with the goal of a final comparison with the gas distribution traced by the [CII] line. In \cite{Romano23}, we modeled the spectral energy distributions of the DGS sources and took advantage of their spectroscopic information to retrieve physical parameters, including stellar masses ($\mathrm{log(M_{*}/M_{\odot})\sim5-10}$) and star formation rates ($\mathrm{SFRs}\sim0.001-10~M_{\odot}~$yr$^{-1}$). We found that most of the galaxies with the highest SFRs show clear signatures of outflowing gas as traced by broad emissions in the high-velocity tails of their [CII] spectra, with the remaining sources likely affected by weaker winds detected in their stacked profile.

\section{Method}\label{sec:method}
When comparing the average spatial distribution of the \textit{Herschel} [CII] and dust continuum with the NUV continuum from GALEX, we wanted to be sensitive to the presence of faint and possibly diffuse emission in the outskirts of our galaxies. Because of this, we performed a stacking of their intensity maps. In the following, we briefly describe the stacking procedure for each tracer, as well as their radial profile extractions.

\subsection{[CII] stacking}
We started from the [CII] intensity maps of the 29 dwarf galaxies of our sample, obtained by summing the spectral channels covering the emission line in the continuum-subtracted PACS data cubes \citep{Romano23}. We ran the Common Astronomy Software Applications package (CASA; \citealt{McMullin07}) task \texttt{IMFIT} to fit a 2D Gaussian function to each map and obtain an initial guess on the corresponding target properties:  peak coordinates, full width at half maximum (FWHM) of the major and minor axis. Because of their noisy intensity maps, the fit did not converge for three sources, which we thus excluded from our analysis. In order not to contaminate the final average [CII] profile of the sample, we also removed three systems that were classified as mergers in \cite{Romano23}, and seven objects with visually disturbed morphology or with the peak of emission shifted  toward the edge of the PACS field of view (which would artificially modify the average emission at the largest spatial scales of our interest). We spatially aligned the remaining 16 galaxies based on their best-fit coordinates, and stacked them through the following equation:
\begin{equation}
    S_{\mathrm{stack}} = \frac{\sum_{i=1}^{N} S_i \cdot w_i}{\sum_{i=1}^{N} w_i}.
    \label{eq:stacking}
\end{equation}
Here $S_i$ is the flux of the generic spatial pixel for the $i$-th galaxy, N is the number of stacked sources, and $w_i=1/\sigma^2_{i}$ is the weighting factor, with $\sigma_i$ the spatial rms associated to each image. As the size of the spatial pixels for each galaxy can be different, we degraded them to the worst spatial resolution among the entire sample (i.e., $3''$) by resampling the images before stacking.

\subsection{Far-IR stacking}\label{subsec:FIR_stack}
To measure the spatial extent of the dust continuum in our galaxies, we took advantage of the \textit{Herschel}/PACS maps at 160~$\mu$m. Observations are available for all but two sources, while no significant emission is detected for 12 galaxies at the position of their optical counterparts. Excluding the latter sources and those flagged as mergers or having disturbed morphology, we ended up with ten galaxies. We made use of the Source Extractor and Photometry tool (SEP) \citep{Bertin96,Barbary16} to estimate and subtract the background from the images, and to detect the presence of objects in the vicinity of the central targets. We then exploited the source parameters retrieved from the SEP (i.e., coordinates, major and minor FWHM, position angle of the modeled ellipse, and peak flux) as a first guess to model and subtract those sources with the PetroFit package \citep{Geda22} to avoid altering the emission profile estimates, and stacked the resulting images through Eq. \ref{eq:stacking}.   

\subsection{Ultraviolet stacking}
We compared the dust distribution with the UV emission from young stars by means of NUV GALEX maps, available for all but five galaxies in our sample. After a first visual inspection of the images, we excluded ten sources with complex morphology (including the three mergers from \citealt{Romano23}) that could hamper the comparison with [CII] and FIR profiles, ending up with 14 galaxies. In the case of multiple detections of the same source, we co-added the images by taking into account the exposure map of each observation. As done in Sect. \ref{subsec:FIR_stack}, we then subtracted the background from each image and removed all the spurious objects in the field of view, before stacking all the galaxies with Eq. \ref{eq:stacking}. Since the FWHM of the GALEX NUV point spread function (PSF) is $\sim5''$, we degraded the resolution of the UV-stacked image to match the one from [CII] and FIR \textit{Herschel} observations (FWHM~$\sim12''$). To do that, we convolved the GALEX image with a 2D Gaussian kernel with $\sigma_{\mathrm{match}}=\sqrt{\sigma_{\mathrm{FIR,PSF}}^2 - \sigma_{\mathrm{UV,PSF}}^2}$, where $\sigma_{\mathrm{FIR,PSF}}$ and $\sigma_{\mathrm{UV,PSF}}$ are obtained by dividing the FIR and UV PSF FWHM by a factor of  2.355.

More information on the galaxies used for the [CII], FIR, and UV stacking and their properties can be found in Appendix \ref{app:stacked_gal}.

\begin{figure}
    \begin{center}
        \includegraphics[width=\columnwidth]{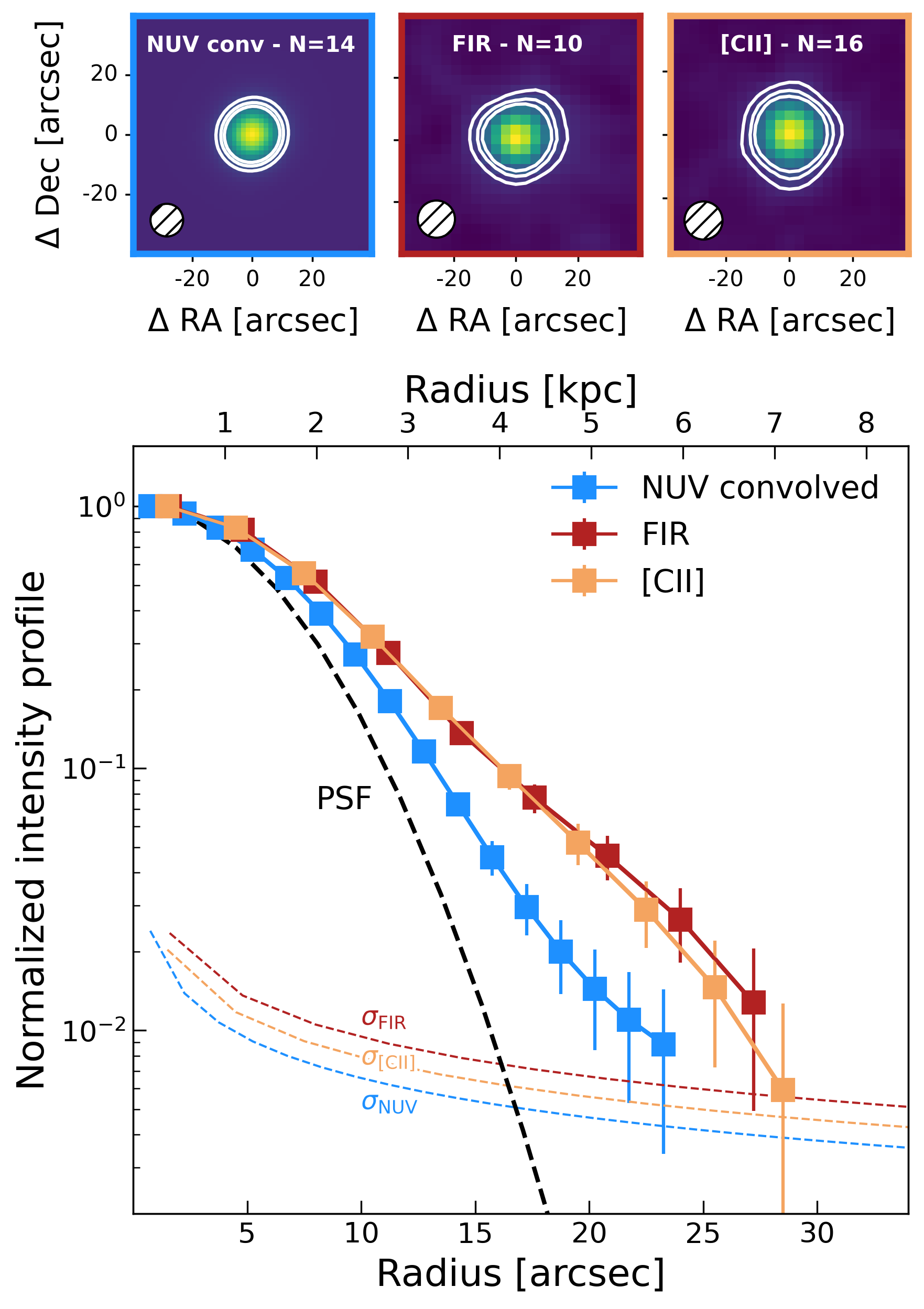}
        \end{center}
    \caption{Stacked images and profiles of the [CII], FIR, and NUV emission. \textit{Top panels:} Stacked intensity maps of the convolved NUV emission, FIR continuum, and [CII] line (from left to right). In each map  the number of stacked sources, and the PSF beam are shown. The contours show the 3, 5, and 7$\sigma$ emission. \textit{Bottom panel:} Normalized intensity profiles for the [CII] line (orange), FIR dust continuum (red), and convolved NUV continuum emission (light blue). The values are obtained in concentric bins of 1 pixel width centered on the peak of the emission. The error bars are computed from the noise maps. The FIR PSF is displayed as a dashed black line. Each profile is shown up to the radius at which the uncertainties reach the corresponding Poissonian noise (colored dashed lines). The distance from the center is shown in arcseconds (bottom axis) and kiloparsec at the mean redshift of the sample (top axis).}
    \label{fig:global_profiles}
\end{figure}

\subsection{Extraction of the radial profiles}\label{subsec:extraction}
For each stacked image we extracted the corresponding circularized radial profile. We centered on the brightest pixel in the image, and built concentric annular apertures of 1 pixel in width. We estimated the mean value for each aperture, dividing it by its area. The uncertainties were obtained by taking into account the total error on each pixel estimated from the background rms noise image, computed through the Photutils package \citep{Bradley23}. For a better comparison, we normalized both the mean values and the errors to the maximum of the corresponding radial profile. The results of this procedure are shown in Fig. \ref{fig:global_profiles}, along with the intensity maps of the stacked images. Since the maps are characterized by different sensitivities, we cut the profiles at the radius at which the uncertainties drop below the corresponding Poissonian\footnote{We estimated the Poissonian noise by dividing the rms of the image by the square root of the number of pixels in each annulus.} noise (see, e.g., \citealt{Ginolfi20b}). 

\begin{table*}
\caption{Estimates of effective radii (in kpc) of the UV, FIR, and [CII] emission by assuming a single and double S\'{e}rsic profile.}
\label{tab:radii}
\begin{center}
\begin{tabular}{c | c | c c c}
\hline
\hline
 & Single S\'{e}rsic & \multicolumn{3}{c}{Double S\'{e}rsic}\\
 &  & central & extended & combined\\ 
\hline
$r_\mathrm{e,UV}$ [kpc] & $1.80\pm0.02$ & $1.68\pm0.02$ (90\%) & $3.61\pm0.19$ (10\%) & $1.94\pm0.03$ (100\%)\\
$r_\mathrm{e,FIR}$ [kpc] & $2.22\pm0.05$ & $1.75\pm0.18$ (76\%) & $3.65\pm0.60$ (24\%) & $2.36\pm0.26$ (100\%)\\
$r_\mathrm{e,[CII]}$ [kpc] & $2.22\pm0.07$ & $1.65\pm0.15$ (62\%) & $3.12\pm0.36$ (38\%) & $2.32\pm0.14$ (100\%)\\
\hline
\end{tabular}
\tablefoot{The value reported in parentheses for the different components of the double S\'{e}rsic profile represents the percentage of light attributed to each component during the fit (i.e., $f_{\mathrm{centr}}$ and $f_{\mathrm{ext}}$ for the central and extended components, and their sum for the combined ones.)}
\end{center}
\end{table*}

\section{Results and discussion}\label{sec:results}
\subsection{Comparison of the radial profiles}
We show in Fig. \ref{fig:global_profiles} the radial profiles for the NUV, [CII], and FIR stacked emissions for our dwarf galaxies. The PSF-convolved NUV emission is found to be more centrally concentrated than the [CII] and FIR emission, with their spatial extension reaching up to 25 and 30 arcsec from the center, respectively (i.e., 6 to 7 kpc at the mean redshift of the samples, $z_\mathrm{mean}=0.012$).

To provide a quantitative characterization of the spatial distribution of the gas, dust, and star formation in our galaxies, we modeled the light profiles with a S\'{e}rsic function \citep{Sersic63}, without making any a priori assumptions on the S\'{e}rsic index. We estimated effective radii $r_\mathrm{e,UV}\sim1.8~$kpc, and $r_\mathrm{e,FIR}=r_\mathrm{e,[CII]}\sim2.2~$kpc, and S\'{e}rsic index $n\sim0.5-0.6$, close to Gaussian profiles. Although the bulk of the intensity for each tracer is well reproduced, the fits are not able to fully recover the emission at the largest radii. Therefore, we adopted a double S\'{e}rsic function, finding a much better agreement with the data than the previous single component (see Appendix \ref{app:models} for more details). In this way, each profile is described by a central and an extended component, with the latter being less significant (in terms of total flux) for the UV emission than the [CII] and FIR continuum. We estimate the effective radius from the combined components in the three profiles as
\begin{equation}
    r_{\mathrm{e,comb}} = \sqrt{f_{\mathrm{centr}}~r^2_{\mathrm{e,centr}} + f_{\mathrm{ext}}~r^2_{\mathrm{e,ext}}} ,
    \label{eq:reff}
\end{equation}
where $r_{\mathrm{e,centr}}$ and $r_{\mathrm{e,ext}}$ are the effective radii of the central and extended component, and $f_{\mathrm{centr}}$ and $f_{\mathrm{ext}}$ represent the fraction of the total flux assigned to each component by the fit, respectively. We obtained $r_\mathrm{e,UV}\sim1.9~$kpc, and $r_\mathrm{e,FIR}=r_\mathrm{e,[CII]}\sim2.4~$kpc. Uncertainties on the radii were estimated by performing a delete-$d$ jackknife resampling \citep{Shao89} consisting of 100 stacking realizations, while removing each time 20\% of the sources in the corresponding sample. For all the stacked images, we performed   the radial profiles extraction again, taking the standard deviation of the resulting distributions as the error. No significant differences were found between the average values of the distributions and those obtained through Eq. \ref{eq:reff}, which suggests that our analysis is not affected by the presence of possible outliers in the samples. The above results are reported in Table \ref{tab:radii}.

\subsection{Comparison with previous works}\label{sec:comparison}
In the past years, many works have found that $r_\mathrm{e,[CII]}>r_\mathrm{e,UV}$ is a widespread condition in high-$z$ star-forming galaxies (SFGs; e.g., \citealt{Carniani18}). For instance, \cite{Fujimoto20} analyzed the physical extent of [CII] emission in a sample of $z\sim5$ SFGs drawn from the ALPINE survey (\citealt{Bethermin20,Faisst20,LeFevre20}). They found that the ratio of the  [CII] to UV sizes typically ranges from $\sim2$ to 3, with a possible increase with   stellar mass, and that $\sim30\%$ of the sample presents extended [CII] halos up to radii of 10~kpc. They claimed that star-formation-driven outflows are most likely  the origin of such halos (see also Sect. \ref{sec:origin}). \cite{Fudamoto22} extended a similar work to $z\sim7$ sources by stacking their [CII] emission from the REBELS survey (\citealt{Bouwens22}). They obtained $r_\mathrm{e,[CII]}/r_\mathrm{e,UV}\gtrsim2$, in agreement with previous results at lower redshift, also suggesting a lack of evolution of the [CII] sizes of SFGs across and beyond the reionization epoch. As for \cite{Fujimoto20}, they attributed the existence of extended [CII] emission to galactic outflows. \cite{Herrera-Camus21} obtained the first evidence of star-formation-driven outflow in an individual $z\sim5$ SFG through dedicated high-resolution [CII] and dust continuum ALMA observations. They found evidence of extended [CII] emission in the source, although with $r_\mathrm{e,[CII]}/r_\mathrm{e,UV}=1.2$, smaller than previous estimations for galaxies at similar redshift. Even so, they claimed the presence of a [CII] halo likely produced by the feedback activity in the galaxy. Finally, \cite{Lambert23} have recently found  strong evidence of diffuse [CII] emission in a $z\sim5.3$ SFG that is two times more extended than the rest-frame UV size. In this case, they suggest a merging event as the origin of the observed extended [CII] emission, with no significant evidence of outflows (although deeper or higher signal-to-noise observations might be necessary for outflow identification).

\begin{figure}
    \begin{center}
        \includegraphics[width=\columnwidth]{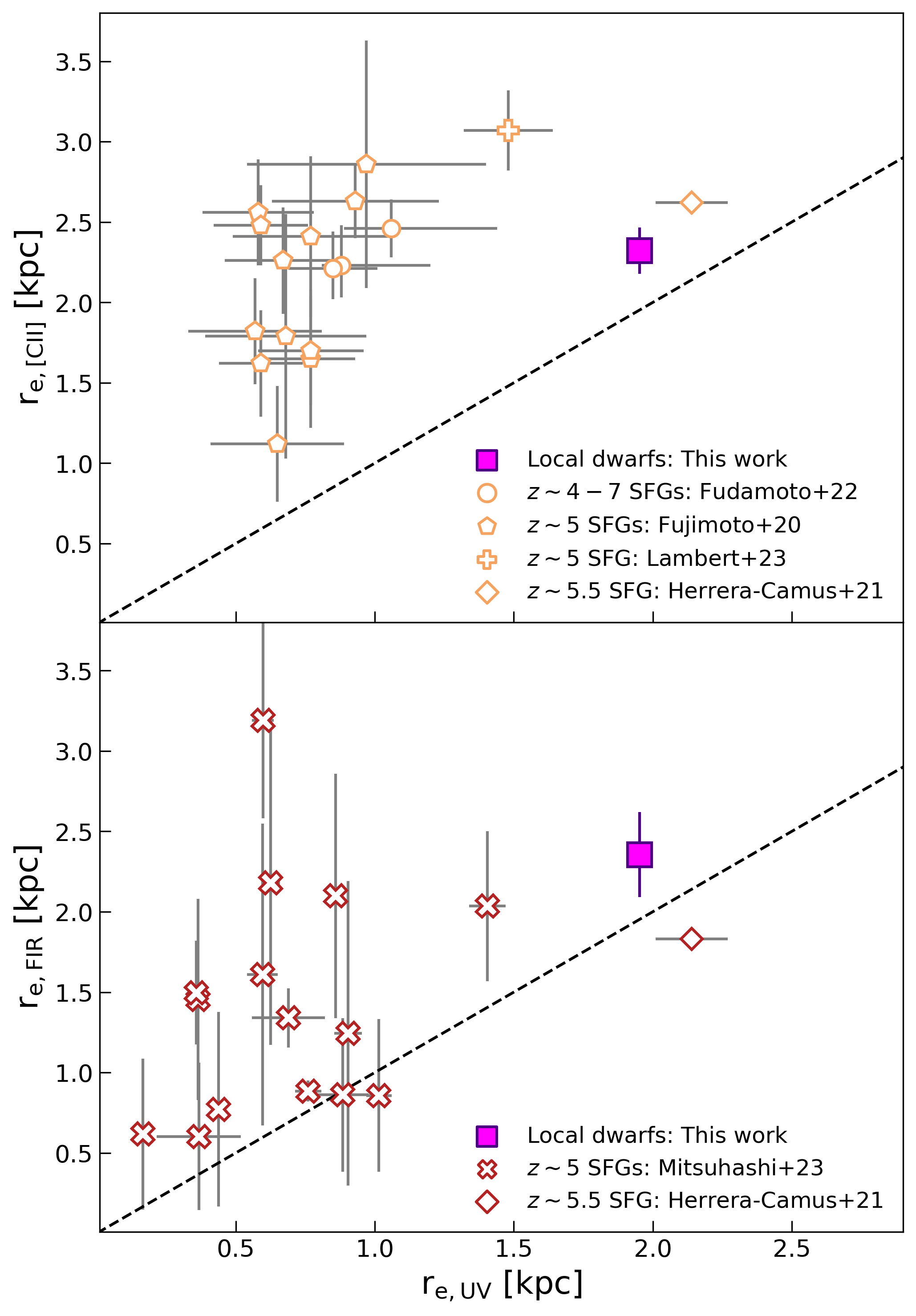}
        \end{center}
    \caption{Comparison of the UV, [CII], and FIR sizes. Empty markers represent [CII] (in orange) and dust (in red) measurements for $z\gtrsim4$ SFGs from the literature (see Sects. \ref{sec:comparison} and \ref{sec:origin}). The magenta squares show the results for local dwarf galaxies from this work. The dashed black lines mark the 1:1 relation.}
    \label{fig:cii_vs_uv}
\end{figure}

We show in Fig. \ref{fig:cii_vs_uv} (upper panel) the comparison between $r_\mathrm{e,[CII]}$ and $r_\mathrm{e,UV}$ for these high-$z$ galaxies\footnote{For ALPINE, we show only the sources with reliable size measurements for both [CII] and UV emission (see \citealt{Fujimoto20}). Also, to be consistent with the other data points, the results by \cite{Herrera-Camus21} were obtained by fitting the [CII], FIR, and UV radial profiles with a S\'{e}rsic function instead of their adopted exponential function.}, and for our local dwarf sources. Our measurements provide a [CII]-to-UV size ratio $\sim1.2$, that is smaller than the average ratio from high-$z$ sources, and in much better agreement with the results by \cite{Herrera-Camus21}. The average [CII] size we obtain is consistent with that from high-$z$ galaxies, possibly suggesting a common mechanism at the origin of the extended emission. In this regard, low-metallicity dwarf galaxies, such as those analyzed in this work, are   considered to be analogs of high-$z$ sources, sharing similar properties in terms of size, (subsolar) metallicity, or specific SFR (e.g., \citealt{Heckman05,Motino21,Shivaei22}). Instead, the UV emission  is  more extended for our local galaxies, in line with a possible $r_\mathrm{e,UV}$ evolution with redshift (e.g., \citealt{Shibuya15}, although this is disfavored by the large UV size measured by \citealt{Herrera-Camus21}). As suggested by \cite{Fudamoto22}, such a differential evolution between [CII] and UV sizes could imply an increasingly large reservoir of gas   compared to star formation activity in galaxies at higher redshift.  

\subsection{Origin of the extended [CII] and far-IR emission}\label{sec:origin}
Following \cite{Fujimoto19,Fujimoto20,DiCesare24}, the formation of the extended [CII] emission around high-$z$ galaxies could be explained by different processes, including satellite sources, large-scale photodissociation (and HII) regions, cold streams, galactic outflows, and tidal stripping. Although all of these phenomena could be responsible for the origin of [CII] halos,   observations and simulations agree on the fundamental role of outflows in the CGM enrichment. 

\cite{Pizzati20} made use of semi-analytical models to simulate the [CII] emission arising from supernova-driven cooling outflows, being successful in reproducing the $\sim10~$kpc [CII] halos observed by \cite{Fujimoto19} in $z\sim6$ SFGs. Additionally, by combining cosmological hydrodynamic simulations and radiative transfer calculations, \cite{Arata20} obtained extended [CII] radial profiles produced by star-formation-driven outflows that also partially matched the \cite{Fujimoto19} data. These theoretical works are supported by observations of cool outflows in the broad wings of [CII] line spectra, in conjunction with evidence of extended carbon emission in the same galaxies \citep{Ginolfi20b,Herrera-Camus21}. For instance, \cite{Ginolfi20b} found outflow signatures in the stacked [CII] spectrum of $z\sim5$ SFGs from ALPINE. When stacking sources with high a SFR, they obtained much more spatially extended [CII] emission with respect to the case of lower-SFR galaxies, highlighting the star-formation-driven nature of the outflow. This could explain the positive correlation between $r_\mathrm{e,[CII]}/r_\mathrm{e,UV}$ and the stellar mass (thus SFR) found by \cite{Fujimoto20}, ultimately suggesting that the physical origin of [CII] halos is linked to galactic outflows. Similarly, in \cite{Romano23} we found that most of the individual local dwarf galaxies hosting prominent galactic outflows have the highest stellar masses and SFRs. Outflow velocities in those galaxies are higher than (or comparable with) the velocities needed by gas and dust to escape their dark matter halos, leading to $\sim40\%$ of the material entrained in the winds   being expelled in the IGM (and possibly an even larger fraction enriching the CGM). We consider all of this information as evidence for stellar feedback producing the extended [CII] emission observed in Fig. \ref{fig:global_profiles}.

In addition, we note that the FIR continuum is as extended as the [CII] emission. This is in contrast with some observational and theoretical works on $z>4$ SFGs usually finding that the cold dust is more compact than [CII] and UV continuum (e.g., \citealt{Fujimoto19,Fujimoto20,Ginolfi20b,Herrera-Camus21,Popping22}). However, a few examples of extended dust emission exist at high redshift. \cite{Fujimoto20} identified one  dust halo object in their sample of ALPINE galaxies, with a FIR continuum profile similar to the [CII] profile. \cite{Akins22} characterized the morphological properties of a strongly lensed galaxy at $z\sim7$, observing an extended dust structure up to $\sim12~$kpc, following the [CII] emission. Recently, \cite{Mitsuhashi23} investigated the spatial extent of dust emission in $z\sim5$ SFGs as part of the CRISTAL survey (2021.1.00280.L; PI: R. Herrera-Camus), which targeted a subsample of the ALPINE sources at higher resolution. They found that the dust continuum can extend up to several kiloparsec, with the average effective radius in agreement with the typical [CII] size of galaxies at that redshift (i.e., $\sim2~$kpc; \citealt{Fujimoto20}), and approximately two times more extended than the rest-frame UV (at odds with expectations from current cosmological simulations; e.g., \citealt{Popping22}). Such findings are also in agreement with our results from the local Universe (see Fig. \ref{fig:cii_vs_uv}, lower panel). This widespread dust in the CGM could be entrained by galactic outflows or tidal stripping of past merging events (e.g., \citealt{McCormick18,Triani20,Kannan21}). In the CGM the dust can rapidly become cold and hard to detect at high redshift due to the low sensitivity and the presence of the warmer cosmic microwave background (CMB; e.g., \citealt{Vallini15,Lagache18}), although correction factors for the CMB can alleviate this problem (e.g., \citealt{DaCunha13}). 

In the local Universe we can reach much deeper sensitivity, while the CMB temperature is low and does not affect the cold dust emission. The stacked FIR profile of our nearby dwarf sources extends to large radii, where dust could have   likely been carried by galactic outflows in the past. This is in line with the results by \cite{McCormick18}, who made use of \textit{Herschel} observations to detect cold dust in the CGM of six local dwarf galaxies affected by outflows. They found that dust can extend well beyond the stellar component, with a typical fraction of $10-20\%$ of the total dust mass residing in the CGM. This fraction increases to $\sim60\%$ for the galaxy with the largest metallicity deficit in their sample (obtained as the difference between the measured metallicity and that predicted by the mass-metallicity relation; \citealt{Tremonti04}), and with the greatest evidence for dust entrained by galactic wind. If the evolution of primordial galaxies is ruled by similar processes as those observed in present-day low-metallicity dwarf sources, we could expect that extended dust components are generally present in the CGM of early SFGs, but too faint to be currently detected. In this scenario the dust mass of high-$z$ galaxies could be higher than estimates not accounting for extended FIR continuum (e.g., \citealt{Akins22}), which has  deep implications for dust production mechanisms at early times (e.g., \citealt{Nanni20,Witstok23}).

\section{Summary and conclusions}\label{sec:conclusions}
We present the first comparison between the spatial emission of [CII]~158~$\mu$m, FIR, and NUV continuum for local, low-metallicity dwarf galaxies. We used the sample of dwarf sources in \cite{Romano23}, taking advantage of GALEX and \textit{Herschel} observations to produce stacked intensity maps in each band. We then compared the average radial profiles of the gas, dust, and stellar activity in the galaxies, and obtained an estimate of their physical size. Our findings are summarized below:
\begin{itemize}
\item The [CII] and dust continuum emissions in our galaxies are more extended than the star formation activity traced by the NUV, which seems to be more centrally concentrated instead. By fitting the radial profiles with double S\'{e}rsic functions, we obtained a $\sim2.4~$kpc radius for both FIR and [CII], and $\sim1.9~$kpc radius for NUV, implying a gas (or dust) to stellar ratio of $\sim1.2$.
\item We compared our measurements with those from $z>4$ SFGs, most of which show diffuse [CII] emission extending up to $10~$kpc. The [CII] size of our sources is consistent with that of primordial galaxies, suggesting a common mechanism at the origin of the emission. On the other hand, the UV continuum appears to be larger, pointing to a differential evolution between gas and star formation activity across cosmic time (e.g., \citealt{Fudamoto22}).
\item  Observations and simulations agree about the fundamental role of galactic outflows in producing the extended [CII] emission detected in high-$z$ galaxies. We found that atomic outflows traced by [CII] broad components are ubiquitous in our sources \citep{Romano23}. In addition, given the resolution and sensitivity we reach in the local Universe, and having excluded galaxies with disturbed morphologies from our analysis, we can rule out any contribution from ongoing merging or satellite galaxies to the observed diffuse [CII] emission in our sample.\footnote{It is worth specifying that mergers and/or satellites can still be responsible for the extended [CII] reservoirs in galaxies' CGM (e.g., \citealt{DiCesare24}), especially in the high-$z$ Universe where interacting systems are more frequent (e.g., \citealt{Duncan19,Romano21}) and observations are hampered by the reduced angular resolution and sensitivity.} All in all, we can thus affirm that outflows are most likely responsible for the CGM carbon enrichment of local dwarf galaxies.
\item As a further imprint of outflow activity in nearby dwarf galaxies, we found that the average dust continuum is as extended as the [CII] emission. Dust in the CGM can  indeed be carried by outflows, as shown by \cite{McCormick18}. At high redshift only a few galaxies show similar extended dust profiles (e.g., \citealt{Fujimoto20,Akins22,Mitsuhashi23}), likely because, in general, it is hard to detect cold and faint CGM dust due to sensitivity limits and the CMB effect. 
\end{itemize}

\noindent
Our results highlight the importance of galactic outflows in the evolution of nearby low-mass galaxies and their primordial counterparts, as well as their imprint on the gas and dust distribution up to kiloparsec scales. Additionally, they could help in calibrating the ongoing [CII] intensity mapping experiments aimed at detecting the faintest SFGs in the Epoch of Reionization, so as to understand their contribution to ionizing photons production and, more   generally, to galaxy evolution in the early Universe (e.g., \citealt{Anderson22,Zhang23}). Observations of individual galaxies at deeper sensitivity (e.g., with the new generation of single-dish telescopes such as AtLAST) will allow us to ultimately disentangle the contribution of different mechanisms (e.g., outflows, minor mergers) to the observed diffuse gas and dust emission in these galaxies.

\begin{acknowledgements}
We thank the anonymous referee for her/his helpful comments and suggestions that improved the quality of our paper. This research made use of Photutils, an Astropy package for detection and photometry of astronomical sources \citep{Bradley23}. This research made use of PetroFit \citep{Geda22}, a package based on Photutils, for calculating Petrosian properties and fitting galaxy light profiles. M.R., A.N., and P.S. acknowledge support from the Narodowe Centrum Nauki (UMO-2020/38/E/ST9/00077). M.R. acknowledges support from the Foundation for Polish Science (FNP) under the program START 063.2023. G.C.J. acknowledges funding from ERC Advanced Grant 789056 ``FirstGalaxies’’ under the European Union’s Horizon 2020 research and innovation programme. J. is grateful for the support from the Polish National Science Centre via grant UMO-2018/30/E/ST9/00082. D.D. acknowledges support from the National Science Center (NCN) grant SONATA (UMO-2020/39/D/ST9/00720).  
\end{acknowledgements}

\bibliographystyle{aa} 
\bibliography{aanda.bib} 

\begin{appendix}

\section{Stacked galaxies}\label{app:stacked_gal}
\begin{figure}[h!]
    \begin{center}
        \includegraphics[width=\columnwidth]{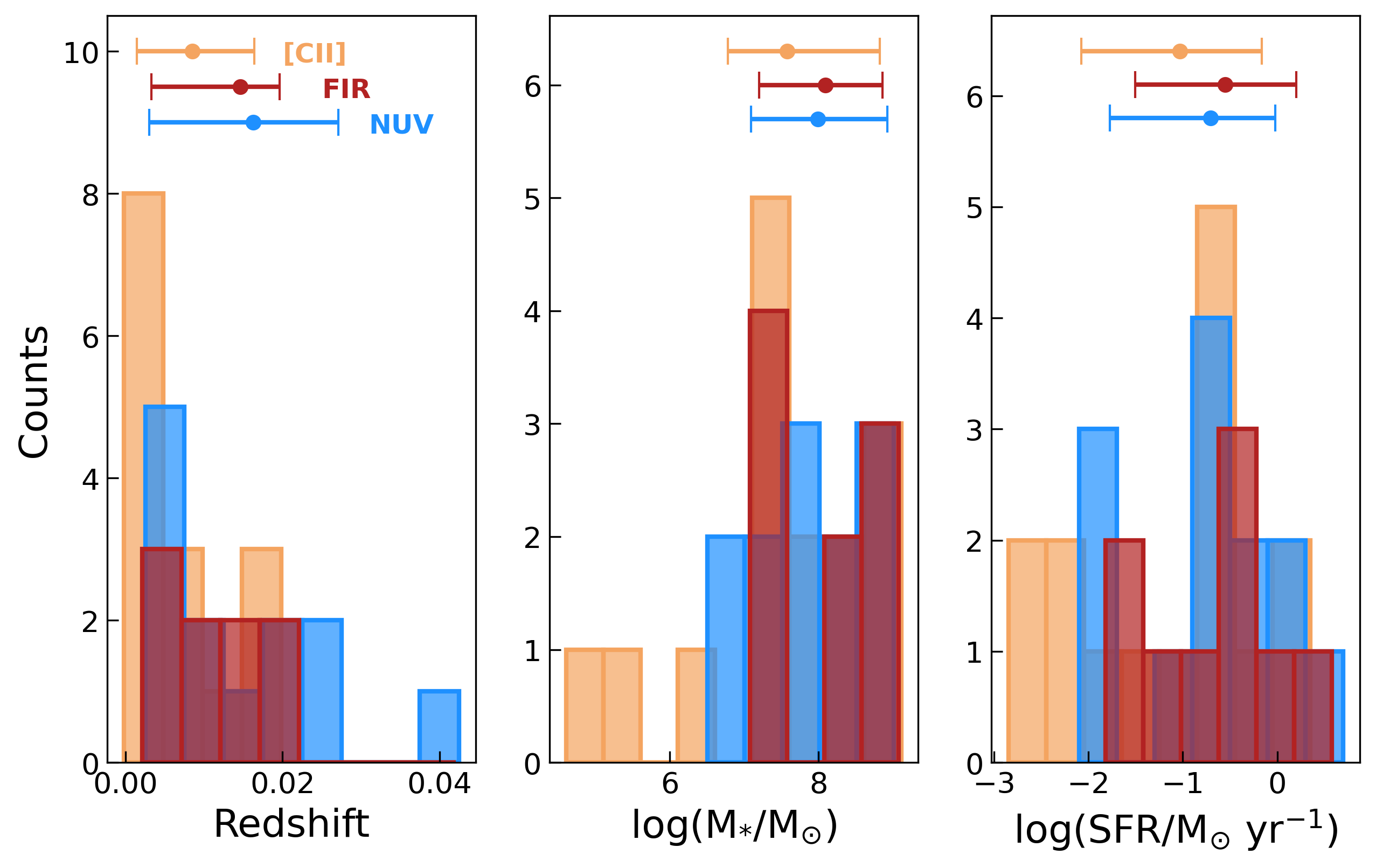}
        \end{center}
    \caption{Redshift, stellar mass, and SFR distributions (from left to right) for our samples of galaxies with [CII], FIR, and NUV emission (orange, red, and light blue histograms, respectively). Circles with error bars (obtained from the 16th and 84th percentiles of each distribution) indicate the average and corresponding uncertainty of the samples.}
    \label{fig:param_distr}
\end{figure}

We list in Table \ref{tab:summary} the dwarf galaxies from \cite{Romano23}, indicating which targets were used in the NUV, FIR, and [CII] stacking. We also show in Fig. \ref{fig:param_distr} some physical parameters of the sources in the final stacked samples, as obtained in \cite{Romano23}. Although our three samples of galaxies have different numbers of sources, they share similar properties in terms of redshift, stellar mass, and SFR. This reassures us that a comparison between the average radial profiles of the stacked samples is still very informative. 

\begin{table}[h!]
\caption{Sample of dwarf galaxies from \cite{Romano23}.}
\label{tab:summary}
\begin{center}
\begin{spacing}{1.25}
\begin{tabular}{l c c c}
\hline
\hline
Source name & NUV & FIR & [CII] \\
\hline
Haro2 & \checkmark & - & \checkmark\\
Haro3 (*) & - & - & -\\
Haro11 & \checkmark & \checkmark & -\\
He2-10 & - & - & \checkmark\\
HS0052+2536 & \checkmark & \checkmark & -\\
HS1222+3741 & \checkmark & - & \checkmark\\ 
HS1304+3529 & - & \checkmark & \checkmark\\ 
HS1330+3651 & - & \checkmark & \checkmark\\ 
HS1442+4250 & - & - & -\\ 
HS2352+2733 & \checkmark & - & -\\ 
IZw18 & \checkmark & - & \checkmark\\ 
IIZw40 & \checkmark & - & \checkmark\\ 
Mrk153 & \checkmark & \checkmark & \checkmark\\ 
Mrk209 & - & - & \checkmark\\ 
Mrk930 & \checkmark & \checkmark & \checkmark\\ 
Mrk1089 (*) & - & - & -\\ 
Mrk1450 & - & \checkmark & \checkmark\\ 
SBS0335-052 & \checkmark & - & -\\ 
SBS1159+545 & - & - & -\\ 
SBS1211+540 & \checkmark & - & -\\ 
SBS1249+493 & \checkmark & - & -\\ 
SBS1415+437 & - & \checkmark & \checkmark\\ 
SBS1533+574 & \checkmark & \checkmark & \checkmark\\ 
UGC4483 & - & - & \checkmark\\ 
UM133 & - & - & \checkmark\\ 
UM311 & - & - & -\\ 
UM448 (*) & - & - & -\\ 
UM461 & \checkmark & \checkmark & -\\ 
VIIZw403 & - & - & \checkmark\\ 
\hline
\end{tabular}
\end{spacing}
\tablefoot{Check marks indicate whether the NUV, FIR, and [CII] maps of the corresponding galaxy have been used in the stacking. Sources flagged with asterisks (*) represent galaxies classified as mergers in \cite{Romano23}, and that have been excluded from the analysis.}
\end{center}
\end{table}

\noindent
\textbf{Galaxy inclination.} In principle, the galaxy orientation could have a significant effect on the stacking results. If galaxies are randomly oriented in the sky, stacking will tend to average over different inclination angles. Instead, in the case of preferred orientation toward edge-on sources, the gas and dust distribution could be biased (e.g., \citealt{Jonsson10}). To check for this, we estimated the inclination angles of our galaxies as $i = \mathrm{arccos}(b/a)$, where $b$ and $a$ are the corresponding minor and major axis from \cite{Madden13}. The resulting distribution of inclinations has an average of $\sim36~$deg, which points toward face-on galaxies (i.e., $i=0$), and that is smaller than the average inclination of 60~deg expected from randomly oriented sources in the sky (e.g., \citealt{Romanowsky12}). Only three galaxies (IIZw40, SBS1415, UGC4483) have inclinations larger than 56~deg (i.e., the 84th percentile of the distribution), thus representing possible outliers in the stacking. However, by removing these sources from the stacking, we found effective radii in agreement within 1$\sigma$ with those obtained from the full samples. Therefore, we can exclude any significant impact of galaxy inclination on our results. 

\section{Radial profile  fitting}\label{app:models}

We show here the comparison between the fit of the NUV, [CII], and FIR radial profiles when assuming a single or double S\'{e}rsic function, as introduced in Sect. \ref{sec:results}. 

\begin{figure*}
    \centering  
    \includegraphics[width=1.3\columnwidth]{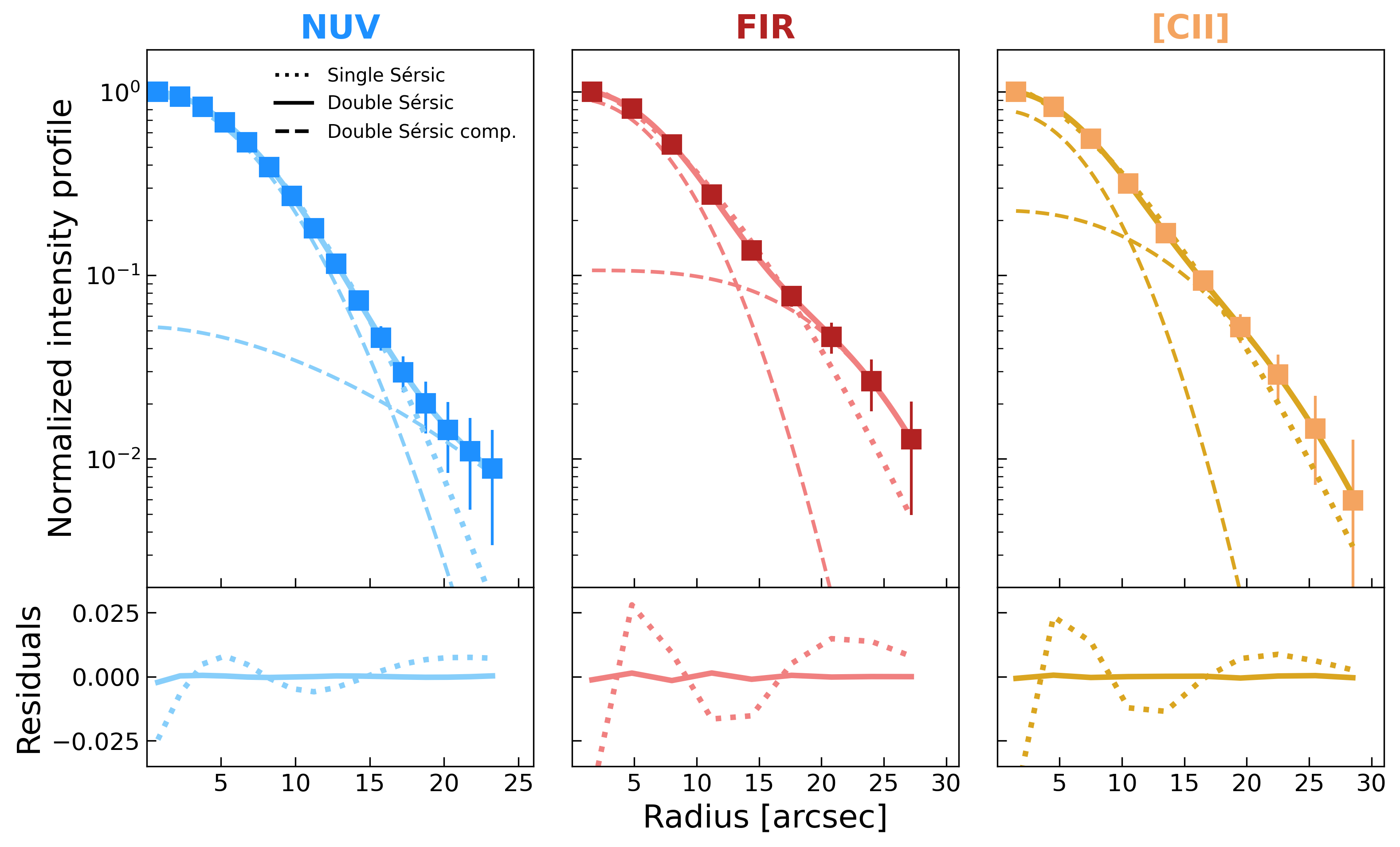}    
    \caption{Stacked profiles for the [CII], NUV, and FIR emission. \textit{Top panels:} Normalized intensity profiles for the convolved NUV continuum, FIR dust emission, and [CII] line (from left to right) represented by the light blue, red, and orange squares, respectively. In each panel the best fit is shown as  a single (dotted line) and a double (solid line) S\'{e}rsic function. In this latter case, the central and extended components of the fit with dashed lines are also shown. \textit{Bottom panels:} Residuals of the fit to the corresponding profile with the single (dotted line) and double (solid line) S\'{e}rsic function.}
    \label{fig:fitting}
\end{figure*}

Figure \ref{fig:fitting} displays the radial profiles extracted as explained in Sect. \ref{subsec:extraction}, along with the residuals obtained with a single (dotted lines) and double (solid lines) S\'{e}rsic fitting. As is evident from the residuals, the latter is in much better agreement with the data with respect to the single S\'{e}rsic, suggesting that a diffuse component is needed to reproduce the profiles. It is worth noting that this extended component is more important for the dust continuum and the [CII] line than for the stellar activity traced by the NUV emission.

\end{appendix}

\end{document}